\documentclass[12pt]{article}

\usepackage{graphicx,anysize}

\begin{document}
\title{``Measurement'' as a neurophysical process: 
\\a hypothetical linear and deterministic scenario}
\author{L. Polley}
\date{\small Institut f\"ur Physik, Universit\"at Oldenburg, 26111 Oldenburg, FRG}
\maketitle
\begin{abstract}\noindent
Tunnel amplitudes of molecular configurations (like neuronal channel pores) 
may be very sensitive to thermal vibrations of the barrier width 
(vibration-assisted tunneling) resulting in pseudo-random spikes 
of widely varying sizes.
An observer who ``lives'' behind the barrier would experience as an 
``event'' an accidental minimum of the barrier width, the timing being
determined by the microstate of the neuron's heat bath. 
In two neurons, set to detect a ``left'' or ``right'' state of an object, 
firing amplitudes typically differ so much as to produce 
a quasi-selection of one option.
\end{abstract}

\tableofcontents

\section{Introduction}

The notion of ``measurement'' as an interruption of deterministic 
evolution governed by the Schr\"odinger equation has proven to be accurate 
and convenient in describing experiments; particularly
so in the framework of open quantum systems \cite{Breuer2002}.
However, it has remained unclear why measuring
devices should not themselves be governed deterministically by the interacting
Hamiltonian of their constituents. Since the 1920s, a peculiar role for 
an observer's consciousness has been suspected by many authors, based on 
von Neumann's observation \cite{vNeumann1932} that the ``collapse of the 
wavefunction'' can be deferred indefinitely by successive measurements, 
terminating only when an observer gets aware of the result.

A partial solution of the measurement problem has been accomplished by 
taking into account the environment-induced decoherence 
\cite{Giulini1996} of macroscopic systems. 
In this approach, a ``pointer'' basis in the Hilbert space of an observing system
is identified which is stable under perturbations by the system's environment. 
When the system evolves according to the Schr\"odinger equation, coupling to an 
object to be observed, a density matrix emerges which is diagonal in the pointer
basis. This amounts to a classical statistical ensemble emerging from a quantum 
state. However, decoherence alone does not provide a mechanism by which the 
environment would determine a particular result of the measurement
\cite{Joos1999,Adler2001}. In fact, such a mechanism is faced with a no-go 
theorem \cite{Ghirardi2000,Gruebl2002} which contends that no 
linear evolution exists which would evolve an arbitrary state of superposition, 
$\alpha|A\rangle+\beta|B\rangle$, 
into a state characterized by property $A$ or property $B$.
The fundamental assumption made in proving the theorem is that the final 
(measured) states are orthogonal, or nearly so at least.   

That latter assumption, however, may not apply to quantum 
brain states of {\em conscious observers}. Brain states fundamental to an 
interpretation of quantum physics have been considered already by a number of 
authors, such as Donald \cite{Donald1990}, Beck and Eccles \cite{BeckEccles1992}, 
Stapp \cite{Stapp1993}, and Mould \cite{Mould2003}. All of those authors assume 
an independent (non-Schr\"odinger) stochastic agent to determine the result of a 
``measurement'',  thus adopting the point of view of open quantum systems. Hence,
there has not yet been an occasion to discuss the mentioned no-go theorem in the 
context of brain states. In the present paper, by contrast, a Schr\"odinger equation (linear 
and deterministic) of a hypothetical mechanism is specified by which microstates
of neuronal heat baths would determine a collective perception of $A$ or $B$ (as
above) by all observers engaged. 
 
As has been stressed already by Donald \cite{Donald1990}, the recent 
progress of the neurosciences \cite{Kandel1995} should encourage discussing
brain states in terms of basic neurophysical notions, such as the firing and 
resting of neurons.  
The clear distinction of firing and resting states
(all-or-nothing principle) can be traced back to molecular origins  
\cite{Kandel1995}. It involves the opening of ion pores, and subsequent passage 
of millions of
ions through the neuronal membrane. Ion pores are protein molecules which 
exist in ``closed'' and ``open'' states.     
Even the opening of a single pore can trigger macroscopic nerval activity 
\cite{Johansson1994}. The hypothesis of the present article is that 
the opening of a pore molecule is a configurational tunnel process, parameterized by
some molecular mass, barrier height and width. The barrier is assumed to be 
thermalized at brain temperature. Aiming at just a proof-of-existence of linear 
and deterministic mechanisms of ``measurement'', 
the potential curve is assumed to be of rectangular shape, with only the 
thickness vibrating thermally.   
      
If that basic notion of the firing and resting of neurons is ``quantized'' 
so as to make Schr\"odinger's equation applicable, it would seem inevitable
to consider an observer's neurons in a state of superposition,
$$ 
  |\mbox{neuron}\rangle = |\mbox{firing}\rangle + |\mbox{resting}\rangle
$$  
where the resting component may be physically large but is {\em not\/},
almost tautologically, {\em part of an observer's experience}. 
Information about the result of a measurement would reside in the firing 
component, so that final states of observers may have considerable overlap in
their resting component. Quantum states of neurons should, in principle, be 
amenable to experimental tests, although a number of decohering effects 
\cite{Tegmark2000} would have to be overcome.  

If the molecular tunneling were stationary, it would show an extreme dependency 
on parameters.
Such a sensitivity does not generally carry over to
time-dependent potentials, as exemplified by periodic time-dependency 
\cite{Hanggi1998}, or by quantum shutters  \cite{Muga2002} in which the 
height of a rectangular potential 
barrier is suddenly reduced from $\infty$ to a finite value.
In the model considered here, the exponential sensitivity is found 
to persist if the {\em width} of the barrier is suddenly reduced to 
a value modulated by a phonon field operator. This here follows by  
analysis of an integral equation; however, the effect has been known as 
{\em vibrationally assisted tunneling\/} since the 1980s when it was
derived for more general potentials by various approximations
\cite{Borgis1989,Benderskii1993}.
When typical molecular parameters are inserted, exponential factors 
$\exp(-\kappa \Phi)$ with vibrational elongations $\Phi$ in the range of 1\AA\  
and $\kappa$ in the range of 30/\AA\ are easily obtained \cite{Borgis1989}.  
In the present application, the sensitivity shows up in 
pseudo-random spikes of tunnel amplitudes, caused by thermal vibrations
encoded in the (nonstationary) quantum state of a heat bath. 
Nearly all of the tunneling is comprised in a narrow interval 
of time, and the sizes of spikes vary over many orders of magnitude. 
Thus, in a state where $A$ and $B$ are entangled with neuron 
$N_A$ and $N_B$, respectively, the numerical coefficient in one summand will 
(almost) vanish relative to the other, the ``choice'' being determined by the
microstates of heat baths $H_A$ and $H_B$.

\section{Quantized thermal vibrations of a barrier width\label{QuantizedVibrations}} 

\subsection{Hamilton operator and integral equation of evolution}

\begin{figure}
\center
\includegraphics[height=55mm,keepaspectratio]{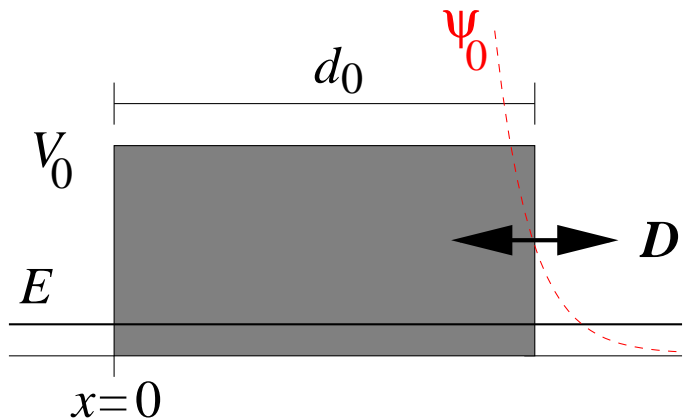}
\caption{Tunnel barrier of fluctuating width $D$ with mean 
value $d_0$. The initial wave function $\psi_0$ of the molecular configuration 
is of the form $e^{-\kappa x}$ with $\kappa$ in the range of
30/\AA.\label{DynBarr}}
\end{figure}
Consider a configuration of effective mass $M$ tunneling under a potential step 
with a right edge in quantized thermal vibrations (Figure \ref{DynBarr}) so that 
the barrier width is 
$$
     D(t) = d_0 + \Phi_0(t)
$$
where $\Phi_0(t)$ is a ``phonon'' field operator (details specified
in (\ref{PhiOp}) below) taken at some ``trigger'' point of the pore molecule. 
Parameters on the left side of the potential are not important for the 
argument. Let us assume the step potential extends to 
$x=\infty$ at times $t<0$ (no tunneling possible).
To emulate a stimulus to the neuron at $t=0$, we assume for $t\geq 0$ 
the Hamiltonian to be
\begin{equation}   \label{HIonField}
    H = \frac{P^2}{2M} + V_0 \Theta(d_0 + \Phi_0 - X) + H_{\rm ph}
\end{equation}  
where $X$ and $P$ are the configurational position and momentum.
The free Hamiltonian of the phonons is
$$
  H_{\rm ph} = \sum\limits_{\vec{k}} \hbar \omega_{\vec{k}} \, 
                   a_{\vec{k}}^* \,  a_{\vec{k}}
$$ 
For $t\leq 0$ the configuration and the phonons are uncorrelated, and the 
configurational wave function $\psi(x,t)$ in the domain of interest, $x>0$, 
is an eigenfunction of $ \frac{P^2}{2M} + V_0$. Assuming an energy level close 
to zero, we thus have at $t=0$
\begin{equation}   \label{psi0}
   \psi(x,0) = \psi_0(x) = C e^{-\kappa x} \mbox{~for~} x>0 
  \qquad\mbox{with~}   
   \kappa \approx \frac1{\hbar} \sqrt{2M V_0}
\end{equation}
$C$ is a normalization factor. The full initial state is
\begin{equation} \label{PsiIn}
   |\mbox{init}\rangle =  \psi_0 \otimes |\mbox{ph}\rangle
\end{equation} 
Let us define 
\begin{equation} \label{Hph+Hpart}
  H_0 = \frac{P^2}{2M} + V_0 + H_{\rm ph} 
\end{equation}
The full evolution operator $U(t)=\exp(-\frac{i}{\hbar}Ht)$ then satisfies the 
integral equation
\begin{equation}   \label{IntEq}
    U(T) = e^{-iH_0T/\hbar}  + \frac{iV_0}{\hbar} \int_0^T 
    U(T-t) \,  \Theta(X - d_0 - \Phi_0) e^{-iH_0t/\hbar} \, {\rm d}t 
\end{equation}

\subsection{Size of the integrand\label{SizeOfIntegrand}}

To evolve the initial state (\ref{PsiIn}) over the interval $[0,T]$ 
using (\ref{IntEq}) we first note that
$$
   e^{-\frac{i}{\hbar}H_0 t}|\mbox{init}\rangle =
 e^{-\frac{i}{\hbar}(V_0-\frac{\kappa^2}{2M})t} \psi_0
 \otimes e^{-\frac{i}{\hbar}H_{\rm ph}t}|\mbox{ph}\rangle
$$ 
Acting with the right-hand side of (\ref{IntEq}) on (\ref{PsiIn}), we aim to 
show that the integrand is strongly peaked at some time $t\in[0,T]$. 
If we measure the size of the integrand by the norm in Hilbert space, 
we can omit phase factors as well as the unitary operator $U(T-t)$, obtaining
$$
  \|\mbox{integrand} \|^2  =  
  \| \Theta(X-d_0-\Phi_0) \psi_0 \otimes  
                  e^{-\frac{i}{\hbar}H_{\rm ph}t}|\mbox{ph}\rangle \|^2
$$
The norm squared involves an integral over the configurational coordinate, 
yielding
\begin{equation} \label{NormSquared(t)}
   e^{-2\kappa d_0} 
 \left( |C|^2 \int_0^\infty e^{-2\kappa x} \,{\rm d}x \right) 
 \langle\mbox{ph}| e^{-2\kappa\Phi_0(t)} |\mbox{ph}\rangle
 \quad\mbox{where}\quad \Phi_0(t) = e^{\frac{i}{\hbar}H_{\rm ph}t}
 \Phi_0 e^{-\frac{i}{\hbar}H_{\rm ph}t}
\end{equation}
The time-dependence of the tunnel matrix element is thus determined by the
free evolution of the phonon field operator in the (non-stationary) 
phononic quantum state $ |\mbox{ph}\rangle$.

\subsection{Heat bath modeled by phonon coherent states}

The heat bath is modeled as a microcanonical phononic ensemble with initial 
conditions set by temporary contact with a reservoir at temperature $T$.
As a workable example of a non-stationary state we assume the phonon field to
be initially in a coherent state.

We consider an $N\times L$ membrane of atoms of mass $m$,
assuming harmonical coupling, periodic boundary conditions, and linear 
dispersion for simplicity. The elongation of an atom at site $\vec{s}
= (x,y)$ is given in the interaction picture by
\begin{equation}  \label{PhiOp}
    \Phi_{\vec{s}}(t) = \sqrt{\frac{\hbar}{2mNL}} 
   \sum_{n=-\frac{N}2}^{\frac{N}2} ~ \sum_{l=-\frac{L}2}^{\frac{L}2}
    (\omega_{\vec{k}})^{-1/2}
    \left( a_{nl} e^{-i\vec{k}\cdot\vec{s}} e^{i\omega t}
  + a_{nl}^* e^{i\vec{k}\cdot\vec{s}} e^{-i\omega t}  \right)
\end{equation}  
where
\begin{equation}  \label{omegak}
  \omega(\vec{k}) = \omega_{\rm D} \sqrt{\frac{n^2}{N^2}+\frac{l^2}{L^2}}
  \qquad \qquad 
   k_x = \frac{\omega_{\rm D}}{c_{\rm S}}\frac{n}{N} 
  \qquad \qquad 
   k_y = \frac{\omega_{\rm D}}{c_{\rm S}}\frac{l}{L} 
\end{equation}  
$a_{nl}$ and $a_{nl}^*$ are phonon annihilation and creation operators 
satisfying canonical commutation relations. The coherent state is represented as
$$
   |\mbox{ph}\rangle = |f\rangle = 
  \exp\sum_{n,l}\left(a_{nl}^*  f_{nl} - a_{nl} f^*_{nl} \right) |0\rangle
$$ 
By standard procedures, the matrix element in expression (\ref{NormSquared(t)}) 
is found to be
\begin{equation} \label{expPhi}
 \langle\mbox{ph}| e^{-2\kappa\Phi_0(t)}|\mbox{ph}\rangle =
 e^{2\kappa^2\langle f, \omega^{-1}f\rangle} \,
 e^{-2\kappa \overline{\Phi_0}(t)}
\end{equation}
where
$
   \langle f, \omega^{-1}f\rangle = \sum_{n,l}\omega_{nl}^{-1}|f_{nl}|^2
$
and
\begin{equation}  \label{PhiClass}
     \overline{\Phi_0}(t) = \sqrt{\frac{2\hbar}{mNL}} 
   \sum_{n=-\frac{N}2}^{\frac{N}2} ~ \sum_{l=-\frac{L}2}^{\frac{L}2}
    (\omega_{\vec{k}})^{-1/2}
  ~ \Re \left( f_{nl} e^{-i\vec{k}\cdot\vec{s}_0} e^{i\omega t} \right)
\end{equation} 
The first factor of (\ref{expPhi}) is time-independent, while the second factor 
carries the time dependence induced by the non-stationary state of the heat 
bath. Since $|f_{nl}|^2$ is the mean phonon number in the $nl$ mode of the 
coherent state, the microcanonical ensemble is approximated here by choosing 
the $f_{nl}$ at random with a statistical weight 
$\exp\left(-\frac{\hbar\omega_{nl}}{k_{\rm B}T}|f_{nl}|^2\right)$
at an assumed temperature of 310\,K. 

\begin{figure}
\hfill\includegraphics{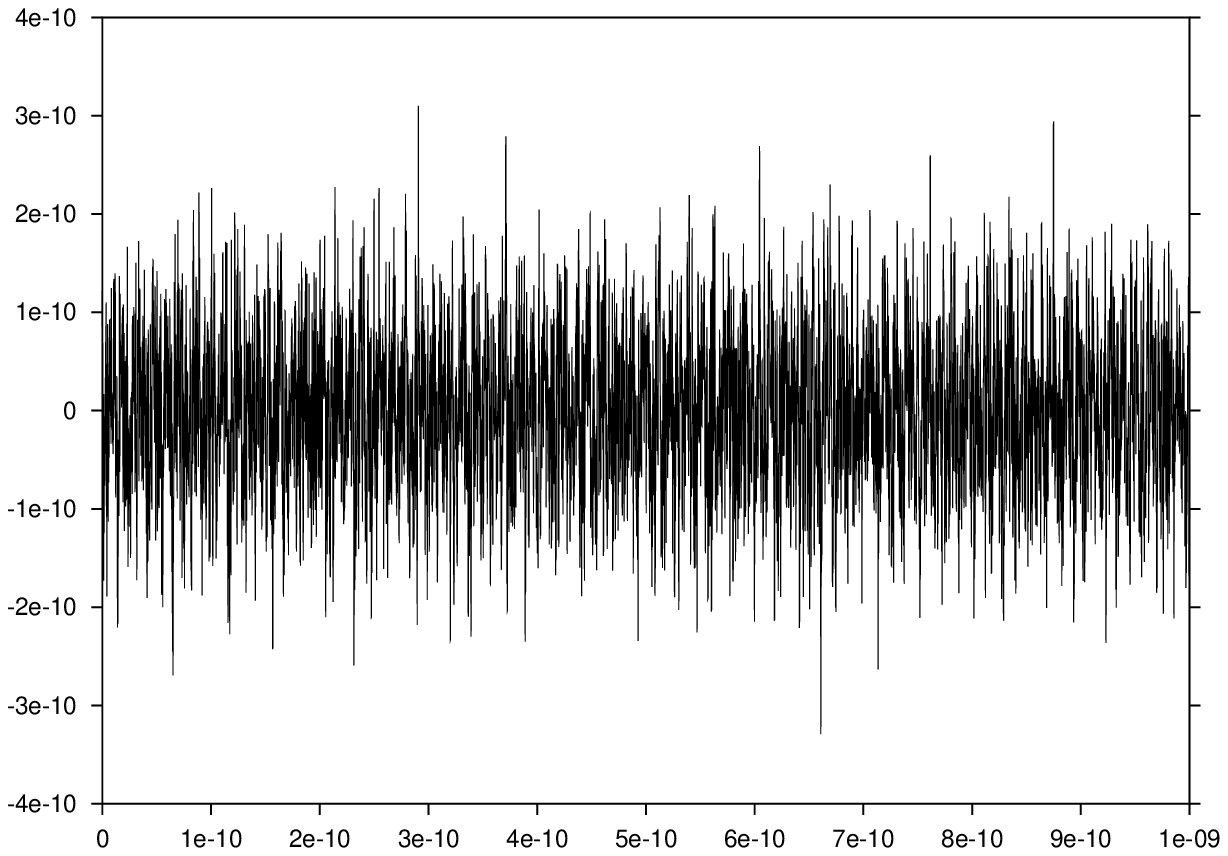}\hspace*{20mm}
\caption{\label{thermfluc} \small Illustration of local thermal vibrations 
$\overline{\Phi_0}(t)$ in an aqueous environment at 310\,K.
Times and elongations are given in seconds and meters.
The corresponding size of the tunneling integrand, as represented by the
factor $e^{ - 2 \kappa \overline{\Phi_0}(t)}$ of equation (\ref{expPhi}) 
with parameter $\kappa$ as in equation (\ref{kappaNumerical}), is shown below.
The lattice size turns out to be an insensitive parameter as to the 
qualitative appearance of the figure; it was chosen here to be $70\times 70$.}
\vspace*{5mm}
\hfill\includegraphics{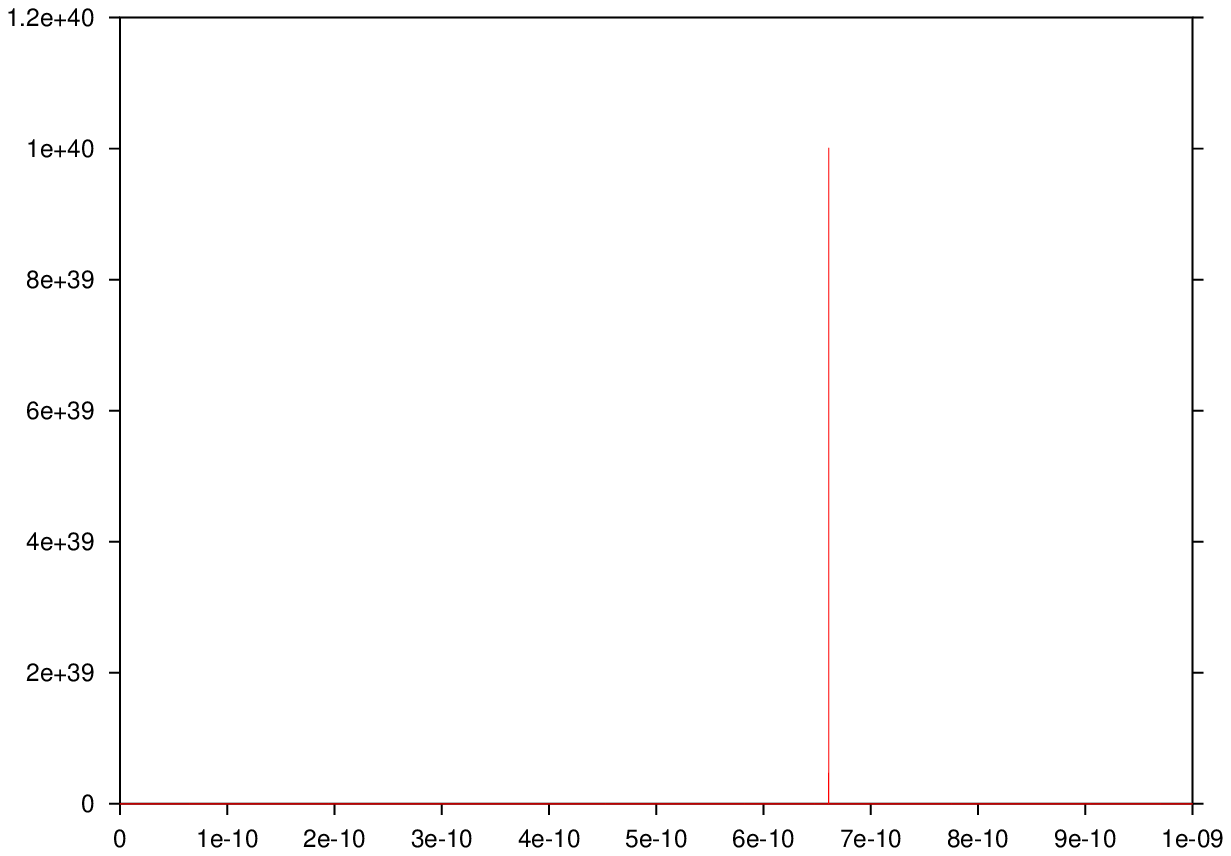}\hspace*{20mm}
\end{figure}

To illustrate the orders of magnitude involved in a range of
parameters as they occur in brains, let us assume a configurational mass of
6\,amu (reduced mass of two carbon atoms) and a barrier height like the 
potential difference across a neuronal membrane, $V_0 = 70\,{\rm meV}$. 
The decay constant of the wave function (\ref{psi0}) then is  
\begin{equation} \label{kappaNumerical}
   \kappa =  14/\mbox{\AA}
\end{equation}  
As to the fluctuations of the barrier width, let us equate them with thermal
vibrations of a water particle (18\,amu) in an aqueous membrane with lattice 
spacing 3\,\AA, speed of sound $c_{\rm S} = 1500\,{\rm m}/{\rm s}$, and 
Debye frequency  $\omega_{\rm D} = 1.6\times 10^{13}{\rm s}^{-1}$.       
A sampling of the time-dependent factor of equation (\ref{expPhi}) 
is shown in Figure \ref{thermfluc}.

\subsection{Full solution: Intuitive arguments\label{FullSol}}

The numerics of Figure \ref{thermfluc} is quite suggestive as to features of a 
full solution of (\ref{IntEq}). Clearly, the peak of the integrand occurs 
at a time $t_{\rm pk}$ when the edge of the barrier is at a position 
$x_{\rm pk}$ of maximal elongation to the left-hand side (see Figure 
\ref{DynBarr}). The wave packet formed under the integral at this instant of 
time, $\Theta(x-x_{\rm pk}) e^{-\kappa x}$, is located entirely to the right of 
the barrier. Subsequently, the edge of the barrier must retract to a less extreme 
position. If it would do so in uniform motion, Galilean symmetry would equate this 
to a wave packet impinging from the right on a stationary barrier, and being 
reflected there. Hence, we expect the wave packet to be {\em expelled\/} from the
retracting barrier. This should be the effect of the operator $U(T,t_{\rm pk})$ 
under the integral of (\ref{IntEq}). Since, with parameters as in Figure 
\ref{thermfluc}, there are no comparable contributions to the integral from 
$t\neq t_{\rm pk}$ we conclude that the norm of the integrand as given by 
(\ref{NormSquared(t)}) and (\ref{expPhi})
already provides a complete picture of the tunneling process.  

\section{A linear mechanism of ``measurement''\label{decision}} 

The main implication of formula (\ref{NormSquared(t)}) is that tunneling
with a large effective mass, when modulated by thermal vibrations, 
can be strongly dependent on the thermal fluctuation that happens to occur. 
This can produce a conscious ``bit'' from a physical ``qubit''.

\subsection{Single observer with two neurons}

\begin{figure}
\center
\raisebox{20mm}{(a)}~~ \includegraphics[height=40mm,width=120mm]{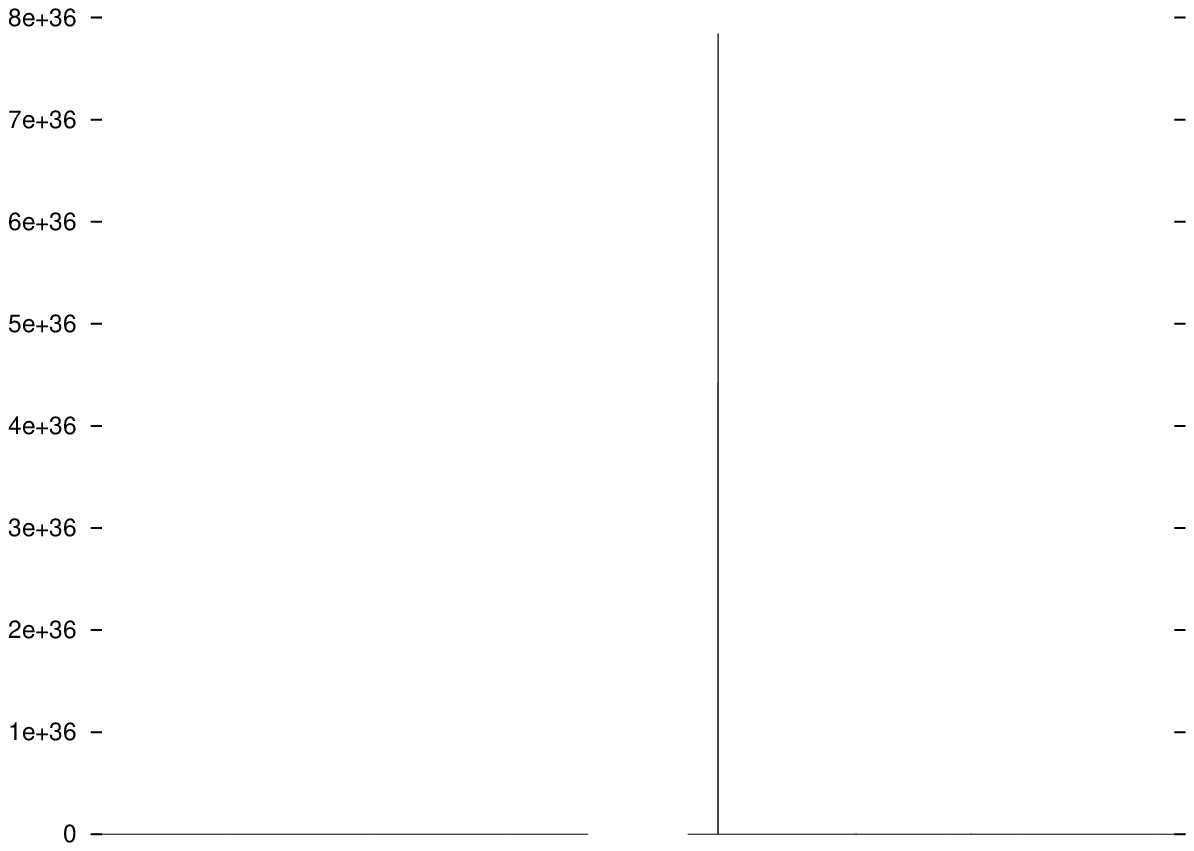}\\[10mm]
\raisebox{20mm}{(b)}~~ \includegraphics[height=40mm,width=120mm]{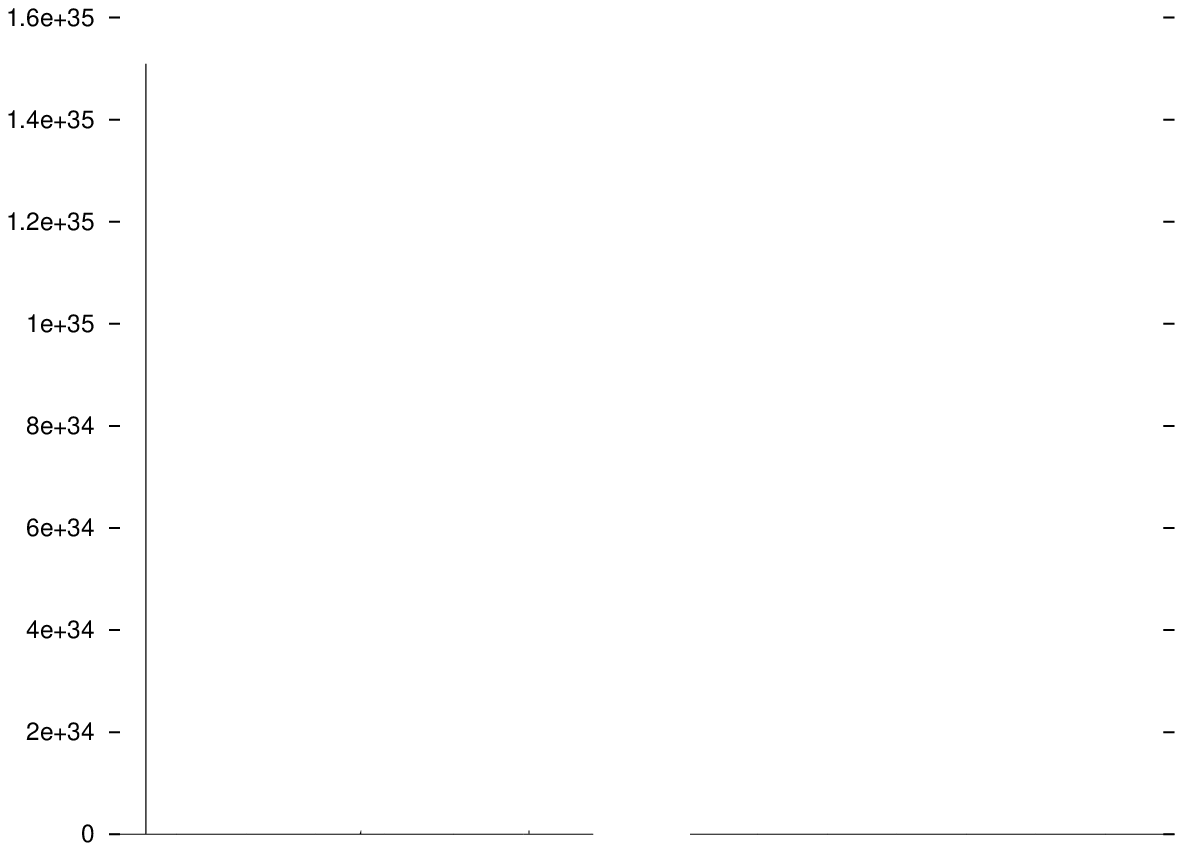}\\[10mm]
\raisebox{20mm}{(c)}~~ \includegraphics[height=40mm,width=120mm]{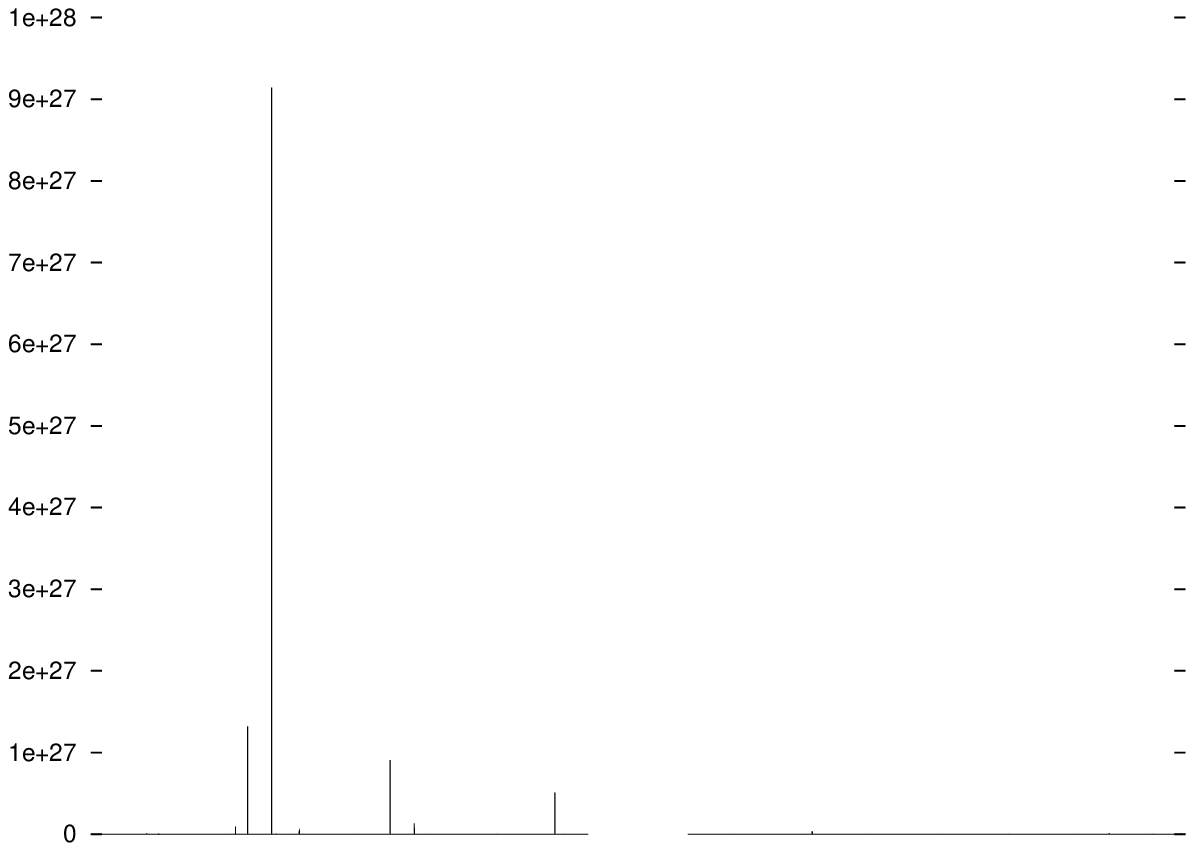}\\[10mm]
\raisebox{20mm}{(d)}~ \includegraphics[height=40mm,width=120mm]{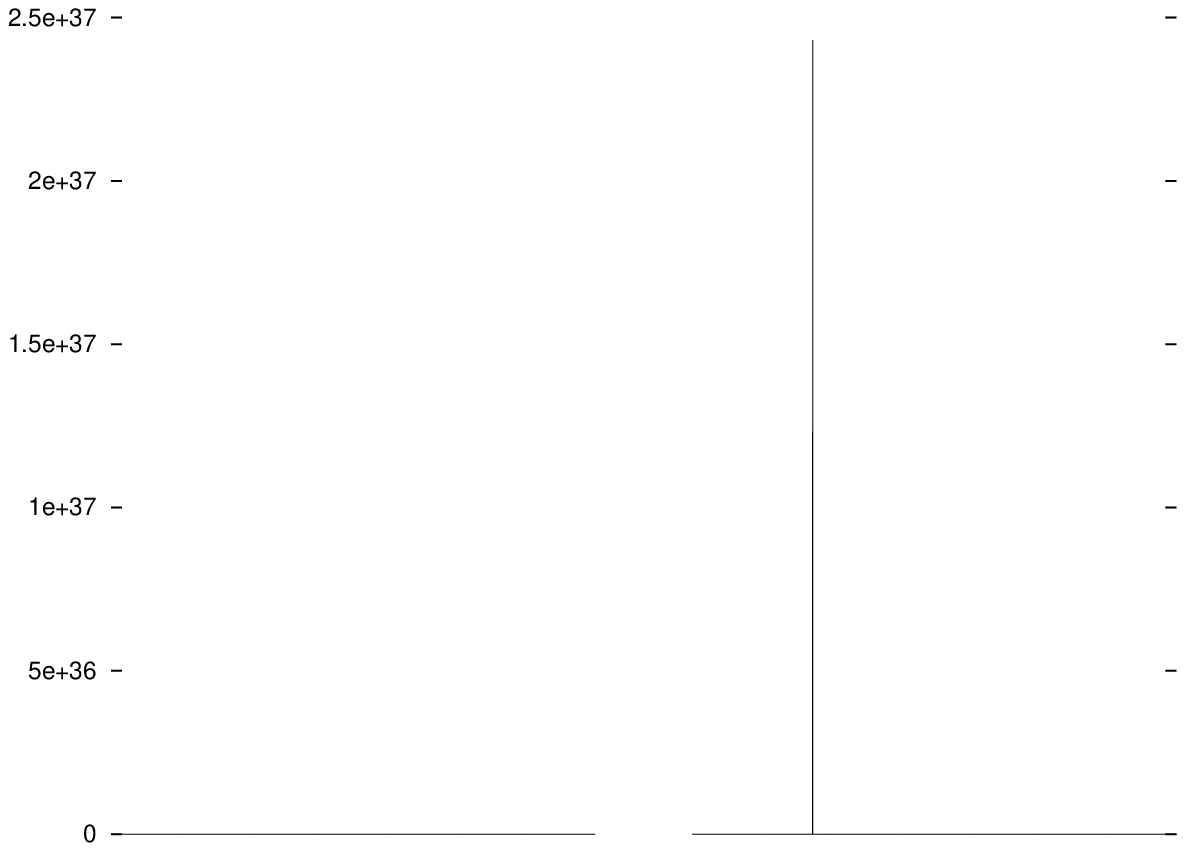}
\caption{\small Four samples of tunnel amplitudes in the {\bf left} and {\bf right}
summand of a superposition, represented as bipartite signal curves.
A {\bf selection} would require one channel to carry a 
zero signal relative to the other. The very different scales of 
the samples should be noted. Multiple spikes as in (c) often
occur if the tunnel amplitude happens to be relatively small.
Also, as more samples would reveal, there is a chance of signals occurring in 
both channels. This ambiguity is reduced if several neurons are involved in 
producing a perception
(cf.\ Section \ref{Objectivity?}). \label{4samples}}
\end{figure}

Consider a superposition of $|\mbox{L}\rangle$ and $|\mbox{R}\rangle$ of some 
quantum object, and a ``measurement'' of this state by two neurons, each of 
which is immersed in its own thermal environment. 
The outcome of ``left'' or ``right'' in the measurement would correspond to the 
firing of one of the neurons.
Let us restrict to the case of equal amplitudes
for ``left'' and ``right'' since more general superpositions can be reduced 
to this \cite{Zurek1998,Deutsch1999}. Assuming the initial states of the neurons 
to be copies of (\ref{PsiIn}) we initially have 
\begin{equation}  \label{initLR}
 \left( |\mbox{L}\rangle + |\mbox{R}\rangle \right) \otimes
 |\mbox{init}\rangle_{\rm L} \otimes |\mbox{init}\rangle_{\rm R}
\end{equation}
where 
$$ 
   |\mbox{init}\rangle_{\rm L,R} =
   \left( \psi_0\right)_{\rm L,R} \otimes |\mbox{ph}\rangle_{\rm L,R} 
$$ 
Following standard theory \cite{vNeumann1932} we assume that,
during the measurement,
the L-component of the superposition will have 
the left neuron evolving nontrivially (its barrier width being finite) and the 
right neuron trivially (its barrier width being infinite). In the R-component, 
the roles of the neurons are interchanged. Thus (\ref{initLR}) evolves into
\begin{equation} \label{entangledLR}
   |\mbox{L}\rangle ~ \otimes
   e^{-\frac{i}{\hbar}H   t}|\mbox{init}\rangle_{\rm L} ~ \otimes
   e^{-\frac{i}{\hbar}H_0 t}|\mbox{init}\rangle_{\rm R} 
\quad + \quad |\mbox{R}\rangle ~ \otimes 
   e^{-\frac{i}{\hbar}H_0 t}|\mbox{init}\rangle_{\rm L} ~ \otimes
   e^{-\frac{i}{\hbar}H   t}|\mbox{init}\rangle_{\rm R} 
\end{equation}
Finally, let us assume that the firing of a neuron requires the configuration
to move  
towards $+\infty$. In a time evolution governed by $H_0$ the potential barrier 
always extends to $+\infty$, so the configuration will never get there. Hence, in the 
L-component of (\ref{entangledLR}) it can only be the left neuron that fires,
and in the R-component it can only be the right neuron. The tunnel amplitudes
of the individual neurons, i.e.\ the sizes of the summands in (\ref{entangledLR}),
are determined by the one-neuron Hamiltonian considered above---and by the 
initial conditions of two {\em independent} heat baths. 
The spikes of the tunnel amplitudes as illustrated in Figure \ref{thermfluc}
will thus not only occur at different times for the individiual neurons, but also 
with largely different sizes within the duration of the measurement.  
Therefore, one of the summands will largely dominate over the other, suggesting 
that a {\em selection} of ``left'' or ``right'' is being experienced. 
To illustrate this feature, a pair of amplitudes produced 
in the same way as in Figure \ref{thermfluc} may be plotted using a common 
scale. Four samples of such pairs of tunnel amplitudes are shown in 
Figure \ref{4samples}.	

\subsection{Approximate statistics of two peaks\label{GumbelApprox}}

In order to obtain analytical estimates,
let us approximate the local phononic elongation 
$\Phi$ (which is a superposition of many harmonic oscillations) 
by a normal distribution with variance $\sigma_0^2$ and zero mean. 
The distribution function of the maxima of $\Phi$ then approaches the Gumbel 
form \cite{Embrechts2003} 
\begin{equation} \label{Gumbel}
 F(\Phi_\mathrm{max}) 
     =\exp\left(-\exp\left(- \frac{\sqrt{2\ln N}}{\sigma_0} 
                  (\Phi_\mathrm{max}-\mu)\right)\right)
\end{equation} 
where $N$ is the (large) number of drawings from the normal distribution, and 
where $\mu$ is a somewhat involved expression comparable in size to $\sqrt{2\ln N}$.
Let us put $\sigma_0/\sqrt{2\ln N} = \sigma$. 

For a ``decision'' between left and right, the peak in 
one channel must be sufficiently large in comparison to the peak in the other. 
Let ``sufficient'' imply a ratio of amplitudes of $e^{\kappa a}$, at least. Thus 
we wish to know the probability for $\Phi_\mathrm{max}$ in one drawing from the 
Gumbel distribution to differ by $a$, at least, from $\Phi_\mathrm{max}$ in the
other drawing. In terms of $\Phi_\mathrm{max}$ in the first drawing, that 
probability is
$$
    \int \Big( F(\Phi_\mathrm{max} - a) + 1 - F(\Phi_\mathrm{max} + a) \Big) 
    \mathrm{d} F(\Phi_\mathrm{max}) 
  = \int_0^1 \left(x^{e^{a/\sigma}} + 1 - x^{e^{-a/\sigma}}\right) \mathrm{d} x 
 = 1 - \tanh \frac{a}{2\sigma}
$$
Similarly, we may consider a more general superposition with unequal coefficients, 
$ \alpha |\mbox{L}\rangle + \beta |\mbox{R}\rangle $,
and ask for the probability that the amplitude of the first term be the larger 
one. This implies $\ln |\alpha| + \kappa \Phi_\mathrm{max,1} 
                 > \ln |\beta| + \kappa \Phi_\mathrm{max,2}$
so that the probability is  
\begin{equation} \label{ProbLR}
    \int F \left( \Phi_\mathrm{max} + \frac1\kappa\,\ln\frac{|\alpha|}{|\beta|} \right) 
    \mathrm{d} F(\Phi_\mathrm{max}) = 
    \frac1{1 + \left|\frac{\beta}{\alpha}\right|^{\frac1{\kappa\sigma}}}
\end{equation}
Since, in favour of a clear ``decision'', $\kappa\sigma$ should be a large 
number, the probability depends rather weakly on the ratio of the amplitudes.
This is a potential problem, as discussed in section \ref{problems}.

\subsection{Several observers and the emergence of objectivity%
          \label{Objectivity?}}

If conscious results of an observation are determined by the vibrations 
of a neuron's heat bath---how can several observers of an object systematically
agree on their results? 

For the question to make sense, all observers must be conscious---a neuron 
must be firing in each observer. 
If there are $n$ observers, each with neurons L and R, the initial state of 
the object-observer system can be written in the same form as equation 
(\ref{initLR}), where now
\begin{equation} \label{InitSeveralObs}
   |\mbox{init}\rangle_{\rm L,R} = \bigotimes_{j=1}^n \left(
   \left( \psi_0\right)_{\rm L,R}  \otimes |j\rangle_{\rm L,R}
   \right) 
\end{equation}
with $|j\rangle_{\rm L,R}$ the quantum state of the thermal environment of 
the left or right neuron of observer $j$.  
Time evolution proceeds as in expression (\ref{entangledLR}) where now
$$
    H = \sum_{j=1}^n H_j
$$ 
Each evolution operator $\exp(- i H_j t / \hbar)$ has an integral 
representation (\ref{IntEq}) of which the interactive part 
produces a conscious state. Hence, objectivity is to be located in the part
of Hilbert space generated by
\begin{equation} \label{ObsProd}
  \left(  \bigotimes_{j=1}^n  \int_0^T
    U_j(T-t_j) \, \,  \Theta(X_j - d_0 - \Phi_{0j}) e^{-iH_{0j}t_j/\hbar}
          {\rm d}t_j   \right)|\mbox{init}\rangle_{\rm L,R}
\end{equation}
where indices L and R refer to the respective summand in superposition 
(\ref{entangledLR}). Due to the product structure of expression (\ref{ObsProd})
there is no physical meaning any more for individual amplitudes of neuronal 
firing---all factors multiply into a collective effect.
 
As in Section \ref{SizeOfIntegrand} we look for strong peaks in the norm of 
the integrand for which we obtain a product of expressions of the form
(\ref{NormSquared(t)}), 
$$
   \|\mbox{integrand}\|^2 = 
   e^{-2n\kappa d_0} 
 \left( |C|^2 \int_0^\infty e^{-2\kappa x} \,{\rm d}x \right)^n 
 \prod_{j=1}^n 
 \langle j | e^{-2\kappa\Phi_0(t_j)} | j \rangle
$$
Evaluating this as in (\ref{expPhi}) we obtain, in particular, the time-dependent
factor
\begin{equation}  \label{expSumPhi}
    \exp\left( -2\kappa \sum_{j=1}^n \overline{\Phi_{0j}}(t_j) \right)
\end{equation}
with $\overline{\Phi_{0j}}(t_j)$ given by (\ref{PhiClass}) in terms of the 
fluctuation parameters $f_{jnl}$ of the $j$th heat bath.
The dominating contribution to the integral is obtained from the  
extrema of the $\overline{\Phi_{0j}}(t_j)$. These extremal values are themselves 
(pseudo-)random variables, each with a probability distribution characterizing 
a single-observer decision. Since a {\em sum} of them occurs in
the exponent of (\ref{expSumPhi}) we conclude by the central limit theorem 
that the fluctuations in the exponent will roughly increase by a factor of 
$\sqrt{n}$ relative to single observers. Thus, the accidental 
dominance of one channel over the other will be even more sharply pronounced 
than with a single observer. 

\section{Discussion}

\subsection{Are tunnel amplitudes ``big enough''?\label{TooSmall?}}

If observers ``live'' entirely in the tunneling tails of their neuronal wave 
functions, there is probably nothing in their immediate self-experience that 
would enable them to discern whether those tails are big or small. However, 
with a decay constant of the wavefunction as in equation (\ref{kappaNumerical}) 
the prefactor $e^{-\kappa d_0}$ in expressions (\ref{NormSquared(t)}) or 
(\ref{expPhi}) would be irritatingly small unless the mean barrier width $d_0$
is taking values below 1\AA.  

In the scenario considered, $d_0$ is a free parameter.
The only {\em theoretical} constraint is that thermal vibrations must 
never (within the opening time of the neuron's ``shutter'') render the 
dynamical thickness $D(t)$ negative.  
Figure \ref{thermfluc} shows that this constraint, too, would put
$d_0$ in the range of intra-molecular distances.

Moreover, if the neuron's stimulated state were kept for sufficently long, it 
would necessarily happen at some instant of time that thermal fluctuations 
render $D(t)$ formally negative, indicating an occasional absence of any tunnel
barrier at all. The duration $\tau$ of such a peak will be determined by the 
inverse Debye frequency. With (\ref{NormSquared(t)}) reducing in such an 
extreme case 
to the factor $ \left( |C|^2 \int_0^\infty e^{-2\kappa x} \,{\rm d}x \right) $
of order unity, the order of magnitude of the integral in (\ref{IntEq}) can be 
estimated as 
$$
  \frac{V_0\tau}{\hbar} \approx \frac{V_0}{\hbar\omega_{\rm D}} \approx 7
$$ 
where the previous values of $V_0 = 70\,{\rm meV}$ and 
$\omega_{\rm D} = 1.6\times 10^{13}{\rm s}^{-1}$ have been inserted. 
The firing component of the neuronal wave function would thus become comparable
in size to the resting component.

\subsection{Potential problems\label{problems}}
 
The scenario considered only refers to a single act of ``measurement''. 
It is not clear 
how an entire {\bf conscious history} would emerge. Since the ``resting'' part 
of the neuronal wavefunction carries no memory of the observation, that part
would have to ``die out'' somehow. The problem may be related to (and possibly
solved by) the fact that actual consciousness (in humans and animals) is a much 
more complicated phenomenon than what has been supposed in this paper.  
According to Edelman \cite{Edelman1989} perceptions even of  
``primary consciousness'' (as it is presumed to exist in animals) are 
fundamentally dependent on previous (remembered) perceptions.
If it were generally true that the present firing of a consciousness-related
neuron {\em requires\/} the firing of similar neurons in the past, then the 
subspace of present firing would be contained in the subspace of past firing, 
thus constituting the physical correlate of a conscious history.   

For a unique peak of tunneling to emerge from thermal fluctuations as in 
Figure \ref{thermfluc}, the configurational wave function parameter $\kappa$
must not be much smaller than the value assumed in equation (\ref{kappaNumerical}).
Obviously, {\bf details of pore molecules} and of their interaction with the 
neuronal membrane would have to be considered in order to see how realistic 
the numbers are which were chosen here from the right ball park, at best.
Moreover, effects of a {\bf dynamical height} of the barrier would have to be
included.  

The exponential sensitivity of the tunnel amplitude was demonstrated for
a barrier of {\bf rectangular shape}. While there certainly exist 
many other deformations of the barrier which would lead to the same effect, 
all of those would be subject to the rather special constraint that 
their deviation from the initial ($t=0$) shape must be small enough 
to overcome the exponential growth of $\psi_0$ towards negative coordinates 
(Figure \ref{DynBarr}).
It is reassuring to note that a sensitivity similar to (\ref{expPhi}) arises
from completely different arguments in other models of vibrationally assisted 
tunneling \cite{Borgis1989}.

In Figures \ref{thermfluc} and \ref{4samples} the simulation 
extended over about $10^4$ cycles at Debye frequency, or about a nanosecond, 
which is very short in comparison to the tens of milliseconds for which a 
neuron keeps its state of stimulation. To extrapolate the {\bf peak statistics}
to an interval of 10\,ms, let us use again the approximation of section 
\ref{GumbelApprox}; in particular, equation (\ref{Gumbel}).  
Thus, if $N$ is increased from about $10^4$ to 
about $10^{11}$, the width of the fluctuations of the maxima {\em reduces}
by a factor of $0.6$. The narrowing trend with increasing $N$ is uncomfortable 
since the fluctuations of the peaks of $\Phi$ are at the core of the above 
mechanism of selection. Anharmonic couplings of phonons would lead to different
statistics of peaks and might be essential here.

Any quantum-mechanical model of measurement should reproduce {\bf Born's rule} 
for the probability of a particular result. It is not clear how this rule 
emerges from the model considered here. A number of authors have shown, by 
various kinds of argument, that Born's rule is not an independent postulate of 
quantum mechanics but is dictated by the superposition principle already 
\cite{Zurek1998,Deutsch1999,Gleason1957,DeWitt1968,Hartle1968}. 
In particular, Deutsch \cite{Deutsch1999} and Zurek \cite{Zurek1998} have
demonstrated that Born's rule for a general superposition of states follows from
the special case of an equal-amplitude superposition in which the rule reduces 
to a symmetry postulate. In case of equal amplitudes, the 
mechanism considered here is in accord with Born's rule 
(cf.\ equation (\ref{ProbLR})). What remains to be clarified, not only in the 
present context but also generally in \cite{Zurek1998} and \cite{Deutsch1999}, 
is by what physical process the unitary transformation of an arbitrary quantum 
state into an equivalent equal-amplitude superposition should be accomplished. 

\subsection{Concluding remarks\label{conclusion}}

A quantum system has been studied which, although governed by a Schr\"odinger 
equation without stochastic or nonlinear modifications, is able to produce a
conscious selection of ``left'' or ``right'' from a superposition of both.
The key assumption was that individual neurons of observers always are in 
``cat states'', i.e.\ superpositions of firing and resting, with only the firing
component being part of an observer's experience. The model is deterministic,
but has the pseudo-randomness encoded in the microstates of heat baths. 

The findings would seem to partially corroborate and partially modify 
the many-worlds interpretation due to Everett \cite{Everett1973}.
In Section \ref{Objectivity?} it turned out that collective selections 
involving many conscious observers are the most robust;
thus, indeed, it seems to be an entire ``world'' that is being selected. 
But there does not seem to be a need for a ``branching'' of worlds---the model 
shows how always one world should be singled out physically. 
In many-worlds terminology, there is only one ``big'' conscious world 
while any alternative conscious worlds are very ``small''.    
\newpage
\begin{small}

\end{small}

\end{document}